# Boosting Resonant Sensing in Fluids: A Surprising Discovery

Sri Harsha Paladugu[1], Kaustav Roy[1], Anuj Ashok[1], Bibhas Nayak[1], Annapoorni Rangarajan[2], Rudra Pratap[1*]

[1]*Centre for Nanoscience and Engineering, Indian Institute of Science, Bengaluru, India-560012*

[2]*Department of Developmental Biology and Genetics, Indian Institute of Science, Bengaluru, India-560012*

*Corresponding author

**Email:** pratap@iisc.ac.in

**Abstract:**

Micro-mechanical resonators are widely used in modern sensing technology due to their high quality-factor (Q), enabling sensitive detection of various stimuli. However, the performance of these resonators in fluid environments is limited by significant viscous and acoustic radiation losses that reduce their Q. Here, we present a paradigm-shifting discovery that challenges the conventional wisdom of resonant sensing in fluids. We report an optimal volume of fluid over a 2D micro-resonator that increases the Q by up to 1000% compared to that in air. We have conducted precise experiments on piezoelectric, circular, membrane-type micro-resonators of 4 mm diameter fabricated using MEMS technology on silicon-on-insulator (SOI) wafers. The top side of the resonator was filled with different volumes of fluid and its Q was measured through resonance tracking by actuating the resonator with an appropriate voltage. We found the existence of an optimal volume of fluid that maximized the Q. We argue that this phenomenon is a result of a balance between the enhancement of kinetic energy of the resonator due to mass loading of the fluid and the energy dissipation through viscous and acoustic radiation losses in the fluid medium. This remarkable enhancement in Q substantially improves the sensitivity of the resonator, with important implications for diverse applications such as biosensing, chemical detection, and environmental monitoring. Our findings challenge the prevailing understanding of resonant sensing in fluids and open up new avenues for the development of highly sensitive and accurate sensors.

**Introduction:**

Microelectromechanical systems (MEMS) based resonant sensors are used in a variety of sensing applications, including mass sensing[1–3], of biological and chemical moieties[4], fluid density[5–11], fluid concentration[12,13], deep tissue activities[14–20], viscosity sensing[21,22], in-air ranging and communication[23,24], and pressure sensing[25]. MEMS-based transducers have been a promising leap in the field of biosensing due to certain fundamental advantages over conventional methods[26,27]. These advantages include compact size, portability, high sensitivity, rapid responses, and a good signal-to-noise ratio. Typically, both 1-D structures such as micro-cantilevers, and 2-D structures such as circular, diaphragm-type, micro-plates are used as MEMS resonators. Numerous sensing applications require micromechanical resonators to operate in liquid and gaseous environments. For example, Kim et al. have measured the density and viscosity of blood using a 2D piezoelectric micro diaphragm[21]. However, these devices suffer from low Q-factor (and hence lower sensitivity) while operating in fluids. Similarly, cantilever-based structures have generated a lot of interest in biosensing applications due to their technological maturity and simple design[28]. But their structural response is severely damped while operating in liquids due to severe viscous losses that result in poor sensitivity[29]. To compensate for this drop in sensitivity, some researchers have attempted to enhance the Q-factor by operating these resonators in higher modes of vibration[30] or by using signal processing techniques[31] at the expense of increasing the complexity of the back-end electronics. Additionally, some novel techniques have been demonstrated to increase the sensitivity of resonant sensors such as using high frequency nano-optomechanical devices with ultralow mass and a moderate dissipation in liquids[32]. Roy et al.,[33] have shown that when the resonance conditions are limited to only intrinsic factors, lowering the Q-factor increases the signal-to-noise ratio, thereby increasing the sensitivity. Similarly, An et al.,[34] have demonstrated the use of buckling, a nonlinear phenomenon, to enhance sensitivity in cantilever structures. While operating micro cantilever resonators in a fluid, Etchrat et al.[35] have reported a maximum Q-factor of 20, whereas Ayela et al.[36] have reported a Q-factor of 150 for a diaphragm-type resonator in fluids. Diaphragm-based sensors provide distinct advantages over cantilevers due to a large surface area for sensing, lower influence of viscous damping, and lower operational frequencies resulting in simpler electronics. Nevertheless, the dynamic response of the diaphragm-based devices is primarily affected by acoustic radiation losses to the surrounding region, reducing their Q-factor[37,38]. This radiation loss has significant effect on the sensitivity of the device[39]. Hence, there is a need to find a way to significantly reduce viscous and acoustic losses in resonators operating in fluids. The current work shows a technique to enhance the Q-factor of diaphragm-based resonators operating in fluids. We have obtained the highest Q-factor (up to 600) at low operating frequencies of less than 15 kHz in water compared to any previously reported diaphragm-based sensors, and that too without any signal conditioning. We report an unintuitive phenomenon of Q-factor enhancement in fluids, which we support with both simulations and experimental results. Our proposed hypothesis explains the observation, where reduced damping effects due to low fluid volume result in high Q and increased sensitivity of resonators. Additionally, we identify an optimal fluid volume that leads to resonators with the highest Q. We also explore some important factors that influence this optimal fluid volume. Our findings open avenues for the development of more sensitive and accurate biofluid sensors.

**Results:**

**Fabrication of 2D resonators:**

In this study, we have used 2D, circular, diaphragm-type of resonators that are generally called Piezoelectric Micromachined Ultrasound Transducers, or PMUTs. The diaphragm is made of several layers of thin films, generally with silicon as the structural layer and lead zirconate titanate as the active piezoelectric layer (Fig. 1(a)). Figure 1(b) and (c) show a PMUT fixed onto a 3D printed holder. This assembly provides the resonator with two distinct interfaces — fluid on the top (it is actually the backside cavity, but we flip it for our experiments, and hence the cavity shows up on top in Fig. 1(b)), and the air beneath. A PMUT's response in a liquid is measured in two ways after filling its backside cavity with the fluid of interest: (a) optically using a Laser Doppler Vibrometer (MSA, 500, Polytech Inc.), and (b) electrically, using a lock-in amplifier. The fluid is filled into the backside cavity using a volumetric pipette for precise control of the fluid volume.

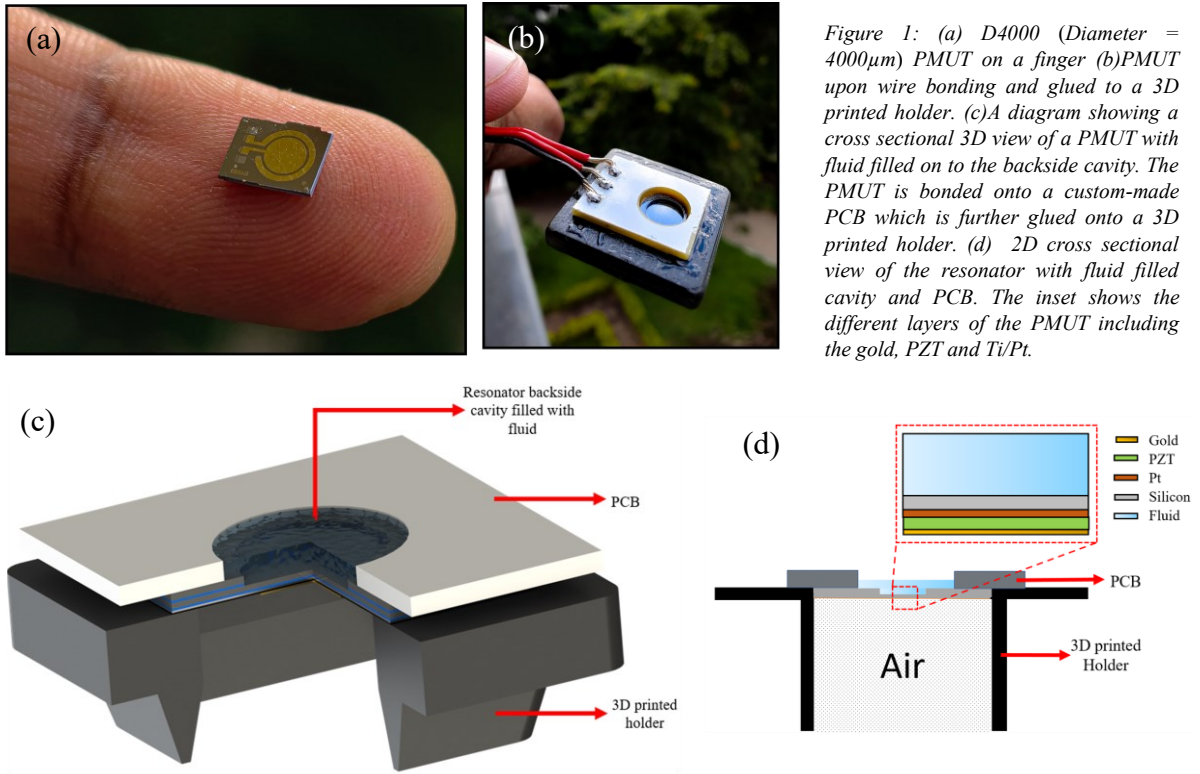

Figure 1: (a) D4000 (Diameter = 4000μm) PMUT on a finger (b)PMUT upon wire bonding and glued to a 3D printed holder. (c)A diagram showing a cross sectional 3D view of a PMUT with fluid filled on to the backside cavity. The PMUT is bonded onto a custom-made PCB which is further glued onto a 3D printed holder. (d) 2D cross sectional view of the resonator with fluid filled cavity and PCB. The inset shows the different layers of the PMUT including the gold, PZT and Ti/Pt.

**Theory of vibration of circular diaphragms in fluids:**

Studies on the vibration of thin plate, or diaphragm-like, structures in liquids were first reported in the late 19$^{th}$ century by Rayleigh[40], and later by Lamb[39], predicting a shift in the natural frequency and damping of such structures. The natural frequency decreases due to the *added mass* of the fluid that vibrates with the membrane. The dynamic interaction of the diaphragm with the fluid also results in viscous losses as well as acoustic radiation losses that lead to higher damping. Lamb's original derivation for the vibration of a clamped circular plate in contact with an inviscid and infinite fluid on one side provides a mathematical prediction for the frequency shift due to the effective increase in mass or the *virtual added mass*. However, we are more interested in understanding the effect of the fluid on damping. The main constituents of damping include (a) energy losses to the surrounding fluid due to acoustic radiation and viscous drag, (b) support losses due to clamping or mounting of the device, and (c) bulk losses due to mechanisms such as thermo-elastic dissipation, piezoelectric damping, and internal friction. The net quality factor, $Q_{total}$, is the reciprocal sum of all the individual corresponding Q's and is given by,

$$\frac{1}{Q_{total}} = \frac{1}{Q_{medium}} + \frac{1}{Q_{support}} + \frac{1}{Q_{bulk}} \qquad (1)$$

Losses associated with the surrounding fluid medium account for most of the energy dissipation for micro-resonators. It includes acoustic radiation and viscous damping losses. The contribution of viscous damping to the Q-factor, as reported by Kozlovsky[41], and can be estimated using the expression,
With

$$Q_{vis} = 2\pi \frac{T_p + T_f}{U_{vis}} = \frac{0.95}{\xi}\left(\frac{1}{\beta} + 1\right) \quad (2)$$

$$\xi = \sqrt{\frac{\eta_k}{\omega_f R^2}} \ \ and \ \ \beta = 0.67 \frac{\rho_f R}{\rho_p h} \quad (3)$$

where $T_p$ and $T_f$ are the kinetic energy of the plate and the fluid, respectively, $\xi$ is a nondimensional parameter that depends on the kinematic viscosity of the fluid, $\eta_k$, the angular frequency of the forcing, $\omega_f$, and the radius of the device, $R$; and $\beta$ is the added virtual mass incremental (AVMI) factor derived by Lamb[39]. The energy, $\alpha$, dissipated through acoustic radiation in the form of sound waves into the fluid can be estimated using Lamb's model, wherein the vibrating structure is considered as a simple acoustic source radiating energy in the form of spherical waves into the medium and is given by,

$$\alpha = \frac{5\pi^2}{9} \frac{\rho_f}{\rho_p} \frac{f_f^2 R^2}{(1+\beta)hC_f} \quad (4)$$

where $\rho_f$ and $\rho_p$ are the densities of the fluid and the membrane, respectively, $C_f$ and $C_p$ are the speed of sound in the fluid and the vibrating plate, respectively, and $R$ and $h$ are the radius and thickness of the membrane, respectively. Hence, the Q-factor for acoustic radiation in the fundamental mode of vibration of the PMUT membrane is given by,

$$Q_{ar} = \frac{\pi f_f}{\alpha} = 1.2 \frac{\rho_p}{\rho_p} \frac{C_f}{C_p} (1+\beta)^{1.5} \quad (5)$$

**Dynamic response of PMUTs in fluids:**

The micro-resonators (PMUTs) were actuated using a 100 mV amplitude AC signal with frequency sweep, and their fundamental flexural modes were characterized using a laser Doppler vibrometer (LDV). Interestingly, the vibration response with ~3 μL water was found to be enhanced ten-fold compared to that in air, which is evident from the computed Q-factor and the membrane displacement (Fig. 2(a)). The membrane displacement in air was 0.36 μm (Q~40), whereas, in water, the obtained displacement was 3.15 μm (Q~450) for the same actuation voltage. This is a 1000 % increase in Q-factor compared to air. To understand this phenomenon, we need to look at the nature of fluid-loading and the boundary conditions during the device operation. Most of the existing analytical solutions for plates/membranes vibrating in fluids assume an infinite volume of the fluid in contact with the resonator resulting in an acoustically radiating boundary condition for the radiated energy. While it is often a valid approximation for macro-scale applications, many MEMS-based sensors working in fluids, have a droplet bounded by air on one side (generally the top) and the device on the other side (generally the bottom), leading to the formation of an air-fluid interface very close to the device surface (with a very small volume of fluid trapped between air and the device). This ensemble behaves differently when vibrating in comparison to the device submerged fully in a large volume of fluid or loaded with a large volume of fluid on one side, as far as energy losses in vibration are concerned. To better understand the differences in loss mechanisms, we calculated the Q-factors using the existing analytical solutions discussed in the previous section. As seen from Table 1, in the fundamental mode, the micro-plate's response is primarily affected by the acoustic radiation losses, evident by huge decrease in Q-factor when acoustic radiation is factored in, both in air and water. The viscous damping effects seem minimal. Weckman and Seshia[37] have also shown that indeed the acoustic radiation is the dominant loss mechanism for a membrane vibrating in a fluid. The Q-factor obtained from experiments (Q~494) is more than the analytically calculated value, even when assuming only viscous losses. This suggests that this mode of operation using a small volume of water on one side of the resonator is able to reduce the acoustic radiation losses significantly, and since the fluid volume is much smaller than the theoretical assumptions (infinite fluid volume), the viscous losses are also on the lower side. As a combined effect, we observe an increased Q-factor and enhanced vibration amplitudes in fluids. While acoustic radiation might be a significant loss mechanism for a membrane vibrating in a bulk fluid, for a micro-resonator operating in a finite fluid condition, viscous losses are dominant as acoustic radiation losses are minimal. Intrinsic losses such as clamping, thermo-elastic dissipation, and piezoelectric damping contributions to the Q-factor at atmospheric pressures are negligible[42] and have not been considered in this study. To understand the effect of the water level in the cavity above the resonator surface (and thus the water volume) on Q-factor, we performed a simple experiment. A PMUT was glued to the bottom of a glass flask where the water level could be easily changed. First, we increased the water level from a fill-level of the backside cavity to 2 cm above the membrane and observed that the Q-factor decreases as the fluid volume increases, as shown in Fig. 2 (b). As the height of the liquid column increases, the energy dissipated in the form of acoustic radiation increases, and hence we observe a decline in the Q-factor (from Q = 77 at cavity filled to the brim, Q = 8 with 2 cm of water above the membrane). A left shift in the resonant frequency is observed due to the increased added virtual mass[7]. Next, we slowly decreased the water level from the fill-level of the backside cavity and measured the response of the PMUT in real-time. we observed an optimal level of water—a *sweet spot of operation*—corresponding to the maximum displacement and an extremely high Q-factor (a Q-factor of 494 for D4000 and 421 for D4500). We label the frequency at which the Q-factor is maximum as $f_s$, *a special frequency*. The results clearly demonstrate that the Q-factor decreases with both increasing and decreasing fluid volume around $f_s$ (Fig. 2(c)).

The high Q-factors achieved for PMUTs in liquids have direct implications in increasing the sensitivity of the sensor. The shift in frequency of a microcantilever resonator in a liquid due to the addition of an analyte ($\Delta f_r$) (For example, a biomolecule) can be expressed as[43],

$$\Delta f_r \propto \sqrt{1 - \frac{1}{2Q_{liq}^2}} \qquad (6)$$

*Table 1: Comparison of the analytically computed total Q for (0,0) mode of a D4000 device for different losses with the experimentally found Q values.*

| Sources of Q | $Q_{total, air}$ | $Q_{total, water}$ |
|---|---|---|
| Viscous losses only | 190.58 | 473.87 |
| Acoustic radiation only | 99.73 | 30.39 |
| Acoustic radiation & Viscous losses | 64.20 | 28.55 |
| Experimental results | 40.5 | 494.05 |

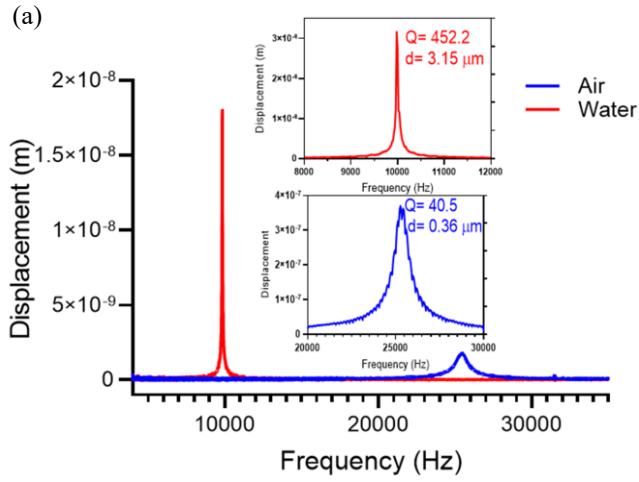
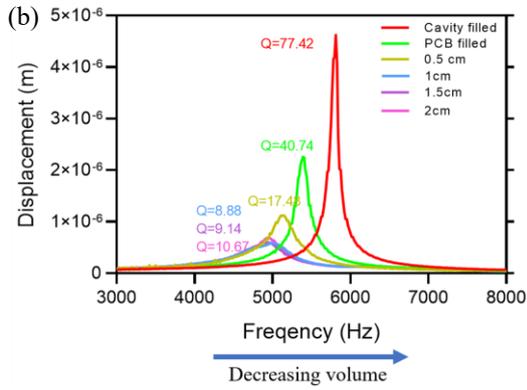
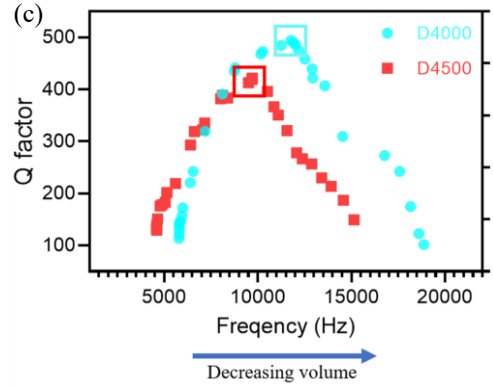

Figure 2: Experimental frequency response of a PMUT with different levels of water. (a) Frequency response of a D4000 PMUT in air and water. The inset pictures show a close sweep of the frequency around the resonance. It also contains information regarding displacement and Q factor in air and water. (b) Shift in frequency and reduction in Q factor can be observed as the height of the water column increases. (c) Measurement of Q factor while reducing the fluid in the PMUT cavity. It resulted in Q-enhancement. The Q factor decreases with increase and decrease of the fluid level around the $f_s$. The coloured big squares contain the optimal fluid loading conditions.

where, $Q_{liq}$ is the resultant Q-factor of the resonator operating in the liquid. Similarly, the mass sensitivity of the resonator can be defined as,

$$S \propto \frac{\Delta f_r}{\Delta analyte} \quad (7)$$

where, $S$ is the sensitivity and $\Delta analyte$ is the change in the analyte which results in the change in frequency ($\Delta f_r$). While the above equations have been derived for microcantilevers, they are applicable for diaphragm-based resonators as well (the proportionality does not change but other terms related to stiffness change). Higher Q increases the sensitivity of the resonator, as evident from eqn. (7), for a given change in the analyte. Specifically, for resonators operating in fluids, which generally suffer from low Q, increasing the sensitivity is of paramount importance. Additionally, the limit of detection (LOD) is also in part decided by the Q-factor. The LOD can be defined as,

$$LOD \propto \frac{\Delta f_{min}}{S} \quad (8)$$

The $\Delta f_{min}$ is the noise-dependent fluctuation of the resonant frequency, which is, in turn, dependent on Q. Lower noise in the system can be observed with a high Q-factor[43]. Lowering noise in combination with increasing sensitivity, increases the LOD of the sensor. Thus, having a sensor with a high Q-factor enhances its sensing capabilities in many ways, and is, therefore, a much-desired attribute.

**Finite element analysis for the Q-enhancement phenomenon:**

To understand the multi-fold enhancement in the Q-factor with a finite amount of fluid on one side, a commercial finite element analysis (FEA) software (COMSOL Multiphysics 5.5) was employed for simulations. The main parameters used to build the 2D resonator model in COMSOL Multiphysics consist of,

- Diaphragm diameter: $D_a$ = 4000 μm
- Diaphragm thickness: $h$ = 25 μm (silicon device layer) + 900 nm (PZT layer)
- Diaphragm density: $\rho$ = 2330 kg/cm$^3$
- PZT pre-stress: $\tau$ = 300 MPa
- Fluid density: $\rho_f$ = 1000 kg/cm$^3$

The acoustic-structure interaction between the resonator and the fluid surrounding the diaphragm was modelled using a 2D axi-symmetric finite element model. The vibrating diaphragm was modelled using the *Solid Mechanics* module. To accurately capture the dissipation in micro-resonators, *Pressure Acoustics* and *Thermo Viscous Acoustics* modules were employed. Based on the previous experimental evidence[44], a pre-stress of 300 MPa was applied to the membrane. Water was modelled in the backside cavity of the device, whereas the other side of the membrane was exposed to air (Fig. 3(a)). The water level was varied across different heights with respect to the PMUT cavity height (400 μm). To simulate the device immersed in a semi-infinite volume (air), a hemispherical radiation space of 50 mm radius above the membrane was used, and additionally, an acoustic-impedance matched boundary condition was applied at the outer end of the air sphere, making it acoustically transparent. The vibrational response was simulated in air and with different volumes of water on top of the resonator, and resonant frequencies along with the corresponding membrane displacements were found in each case.

First, we simulated the frequency response of the PMUT in air and water to verify the experimental results discussed in Fig. 2(a). Similar to the experimental results, we observed an enhanced amplitude in water compared to that in air (Fig. 3(b)). This suggests that the finite element model can indeed capture the phenomenon we have experimentally observed. To analyse the role of the fluid mass in the phenomenon, we compared the model's response without water but with increasing density of the membrane. The increased density mimics the fluid mass addition without the accompanying losses in the fluid. The idea was to delineate the effects of mass and loss mechanisms and understand which parameter plays what role. As seen in Supplementary Fig.1 (see Supplementary Material), with the increasing density of the membrane, the frequency shifts leftward as expected (due to increased mass). Additionally, the displacement also increases with the density (actuation voltage kept constant in both cases) due to the enhanced kinetic energy. However, the peak displacement in water is found to be twice as that obtained with an increase in density for an equivalent mass of the membrane (3 μm as compared to 1.5 μm), at the same resonant frequency of 10 kHz (Supplementary Fig.1 and Fig. 3(b)). The comparison suggests that the fluid mass is essential, but the fluid properties that dictate the loss mechanism also play a significant role in Q enhancement. To further understand the role of the fluid, we varied the fluid volume inside the PMUT cavity. Interestingly, as the fluid volume decreases in the cavity, the peak displacement at resonance increases initially, reaches a maximum, and then decreases (Fig. 3(c)) as depicted by experimental results in Fig. 2c.

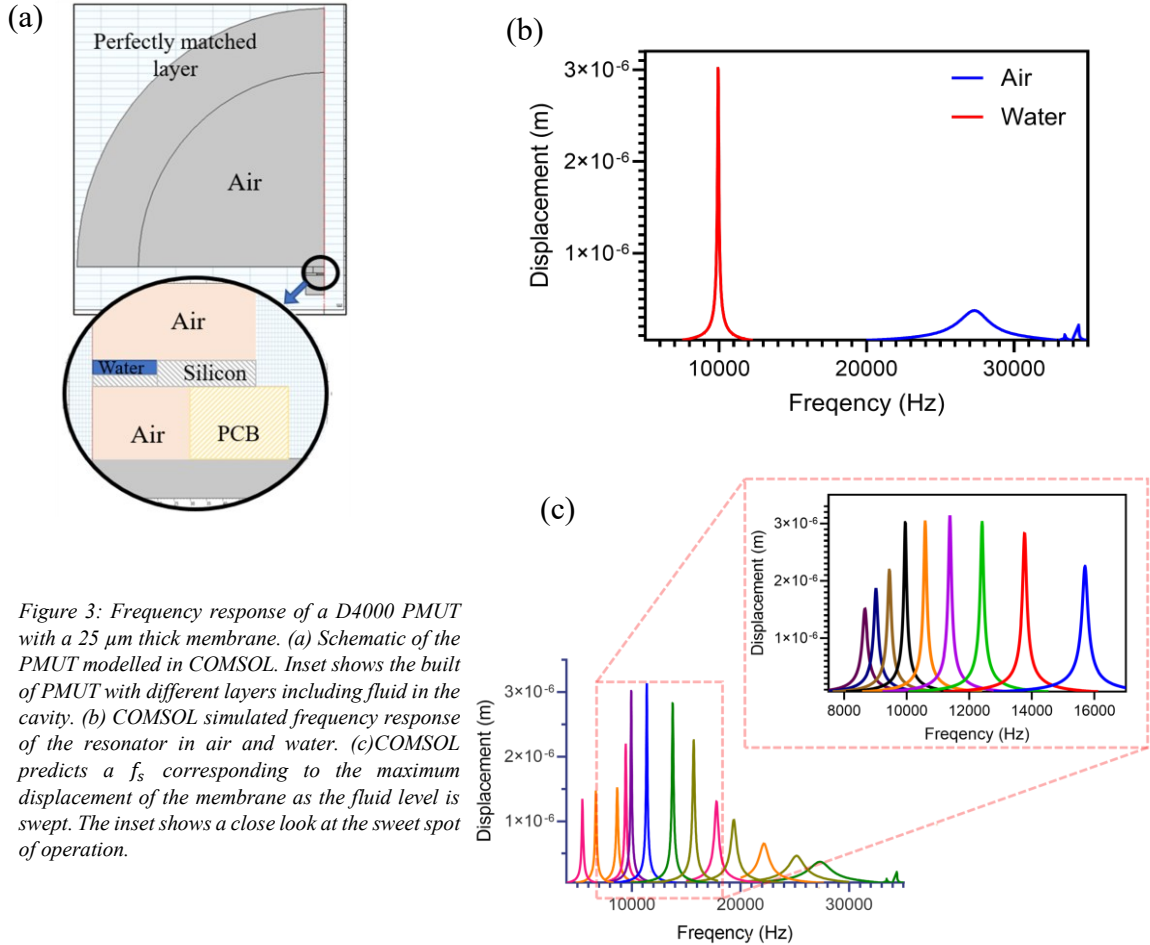

*Figure 3: Frequency response of a D4000 PMUT with a 25 μm thick membrane. (a) Schematic of the PMUT modelled in COMSOL. Inset shows the built of PMUT with different layers including fluid in the cavity. (b) COMSOL simulated frequency response of the resonator in air and water. (c) COMSOL predicts a $f_s$ corresponding to the maximum displacement of the membrane as the fluid level is swept. The inset shows a close look at the sweet spot of operation.*

To further understand the Q-enhancement phenomenon, we developed an equivalent single-degree-of-freedom (SDOF) mass-spring-damper model incorporating the data (frequency, displacement, kinetic energy, and Q-factor) from COMSOL simulations. First, we extracted the mass and stiffness values of the SDOF model corresponding to different levels of water inside the cavity. This was done by equating the maximum kinetic energy obtained from COMSOL simulations to the maximum analytical kinetic energy of the model ($E_k = \frac{1}{2}m\omega^2 x^2$), where $\omega$ is the natural frequency of the fluid-loaded membrane and $x$ is the maximum displacement at resonance. Next, the dynamic response curves around each resonant frequency corresponding to the different levels of water, as shown in Fig. 3(c), were obtained from the analytical solution of the SDOF system's equation of motion, incorporating the corresponding mass and stiffness values extracted from the COMSOL simulations but with a constant damping ratio of $\xi = 0.01$ (Supplementary Fig. 2). The idea behind generating these results and comparing them with Fig. 3(c) was to see if the enhanced mass alone could account for such significant enhancement in Q values. So, if the mass were solely responsible for the Q-enhancement, keeping the damping ratio constant would have resulted in a displacement trend similar to that in Fig. 3(c). However, with increasing fluid mass, the peak displacement at resonance is seen to be decreasing (Supplementary Fig. 2) unlike what we see in Fig. 3(c). This shows that the change in mass alone is not the primary cause of Q enhancement. To confirm the role of damping, the COMSOL calculated Q-factors were used to determine the corresponding damping ratio ($x = \frac{1}{2Q}$) for the equivalent SDOF model, and these values, in turn, were used to calculate the damping co-efficient ($C$) and the critical damping ($C_c$). The logic behind this calculation is that if the change in Q occurs due to a change in the fluid damping, the damping co-efficient $C$ should undergo a minimum at $f_s$, the special frequency where Q becomes maximum. As seen in Fig. 4(a), the damping co-efficient ($C$) indeed undergoes a minimum near $f_s$, whereas the critical damping ($C_c$) is on a decreasing trend (Supplementary Fig. 3). This implies that at $f_s$, the damping value is minimum and, therefore, the corresponding displacement amplitude and the Q-factor are maximum. We now turn our attention to understand the fluid volume/height at which this phenomenon occurs. We simulated the dynamic response of PMUTs of three different diameters (D3500, D4000, D4500 μm), each with varying height of water column in the cavity and computed the Q-factor for each response in addition to recording their peak displacement amplitude. Figure 4(b) shows the computed Q values (left y-axis) against the PMUT's excitation frequency for each of the three devices, along with the fluid-height to device-diameter ratio, H/D, on the right y-axis. We noted the special frequency $f_s$ that corresponds to the maximum Q-factor and extracted the *optimal* fluid-height from the corresponding H/D value. Interestingly, all the optimal H/D ratios lie in a very close range (0.026-0.035 or, approximately 3/100) indicating that there is a critical value of H/D that results into the maximum Q. Using this data for three different H/D ratios, we plot a graph in Supplementary Fig. 4 that predicts the volume of water for the maximum Q-fator for a resonator with arbitary diameter.

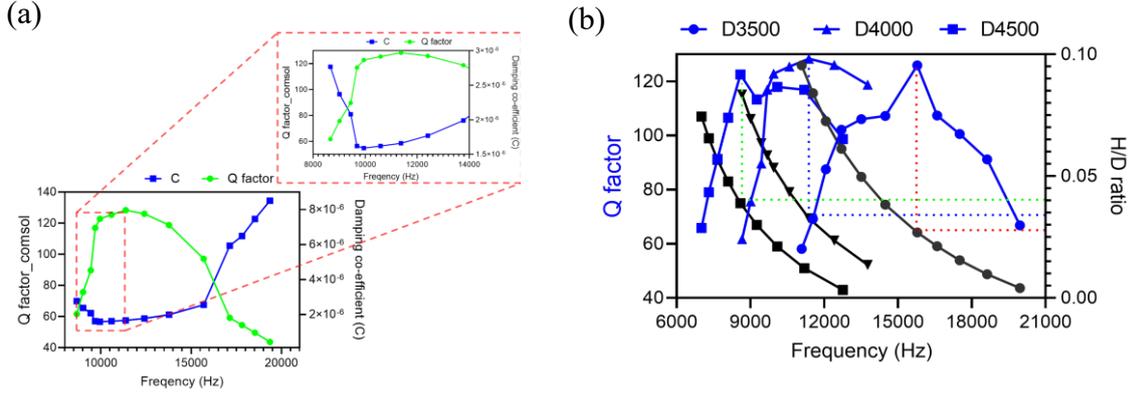

*Figure 4: COMSOL Simulations of PMUTs with fluid investigating the relationship between fluid level and response of the device. (a) The single DOF model of the resonator with fluid is shown. As the Q-factor goes through a maximum, the damping co-efficient (C) goes through a minimum. (b) The Q-factor (left y-axis) and the fluid-height to device-diameter ratio, H/D, (right y-axis) vs the vibration frequency. The H/D ratio corresponding to the maximum Q-factor frequency $f_s$ lies in a very small range (0.026-0.035) and decreases with diameter.*

**Hypothesis and experimental evidence for Q-enhancement:**

To explain this unusual Q-enhancement of the micro resonators with a certain optimal volume of fluid, we propose a hypothesis. As seen in Fig. 5(a-d), we assume that a certain fluid volume (X) moves with the membrane as a rigid mass as the membrane vibrates (we call it *displacement volume*), whereas the rest of the fluid (H-X) acts like a sink to which a fraction of the vibrational energy is dissipated. It is important to note that there are two sinks for energy here: (1) a part of the vibrational energy is lost to the outside air environment above the fluid column and into the air cavity below the membrane in the form of acoustic energy, and (2) a part of the vibrational energy is lost to the fluid column in the form of viscous losses and acoustic radiation. The fractional energy lost to the fluid column depends on the volume of the fluid as we explain here. In Fig. 5(a)-(d), the red colour indicates the vibrating Si membrane, the blue colour indicates the total fluid column on top of the membrane (H), and the yellow represents the threshold fluid column (X) that moves rigidly with the membrane. Figure 5(a) represents the case where the total fluid height is greater than the rigidly vibrating fluid (H>X). Here, the yellow fluid column losses a part of its energy to the blue column that acts as a sink. As the height of the sink decreases (Fig. 5(b)), say, from H to H', where H'<H, the amount of energy absorbed by the sink goes down. In comparison, the amount of energy lost in Fig. 5(a) is greater than in Fig. 5(b), and hence the Q-factor increases from the case in 5(a) to that in 5(b). This phenomenon continues as the fluid volume in the cavity keeps decreasing, thereby increasing the Q-factor. We call this region the *loss dominated region*. Once the fluid level reaches that of the rigidly vibrating fluid column X (Fig. 5(c)), the amount of energy lost to the sink is minimum and thus this case results in the maximum Q-factor. Further, as the fluid volume drops below X (Fig. 5(d)), the entire fluid column still vibrates rigidly along with the membrane, similar to when the volume was X, however, the external losses become constant (as the fluid column along with the membrane radiates acoustic energy to air as before), and the kinetic energy of the system decreases on account of the lower fluid mass. The decrease in kinetic energy can be confirmed by back calculating the speed of the system. The sensing voltage (actual measured output from the device) is a surrogate for the device displacement and follows the same trend as the Q-factor around the $f_s$ (Supplementary Fig. 5(a)). The speed of the device (sensing voltage x frequency) also follows the same trend as displacement, decreasing after $f_s$ (Supplementary Fig. 5(b)). Thus, the speed and the mass of the system decrease and consequently, the kinetic energy is on a downward trend after $f_s$. The combined effect of the acoustic radiation losses remaining constant and the reduction in the kinetic energy leads to a decrease in the Q-factor. This trend continues further as the fluid height decreases and reaches a point where the stability of the fluid thin film starts to play a role. We call this region the *kinetic energy dominated region*. After a critical fluid film thickness, a spontaneous de-wetting occurs due to higher displacements (Supplementary Fig. 6) at the centre of the membrane. So, we propose that the initial reduction in the fluid volume results in a reduction in the viscous and acoustic radiation losses, which is the primary reason for the observed enhancement in the Q-factor. After the special frequency $f_s$, the reduction in the Q-factor is mainly due to the drop in kinetic energy on account of the decreasing fluid mass that vibrates rigidly with the membrane (Fig. 5(e)). While this is a simplified model of the underlying phenomenon, the actual profile of the fluid column (X) and the total fluid height (H) in Fig. 5(a)-(d) may differ slightly because the evaporation surface profile of the fluid follows a parabolic shape tapering towards the centre due to the surface tension forces between the fluid and the cavity wall. To corroborate this hypothesis, we actuated the devices at multiple voltages ranging from 10 mV to 30 mV and recorded the dynamic response as the fluid height decreased inside the cavity. As seen in Fig. 5(f), in the loss-dominated region, the responses overlap for each actuation voltage, suggesting that as long as the fluid level is greater than the threshold value (X), the Q-factor depends on the fluid volume and not the actuation voltage. However, as the fluid level reaches a threshold value and decreases further into the kinetic energy dominated region, increasing actuation voltage imparts higher speeds to the membrane, thereby boosting the kinetic energy and, consequently, increasing the Q-factor with the increasing voltage. This resulted in a maximum Q-factor of 602 in a D4000 PMUT operating at 30 mV actuation voltage. As far as we know, this is the highest Q-factor ever reported for a micro resonator operating in water in the fundamental mode of vibration.

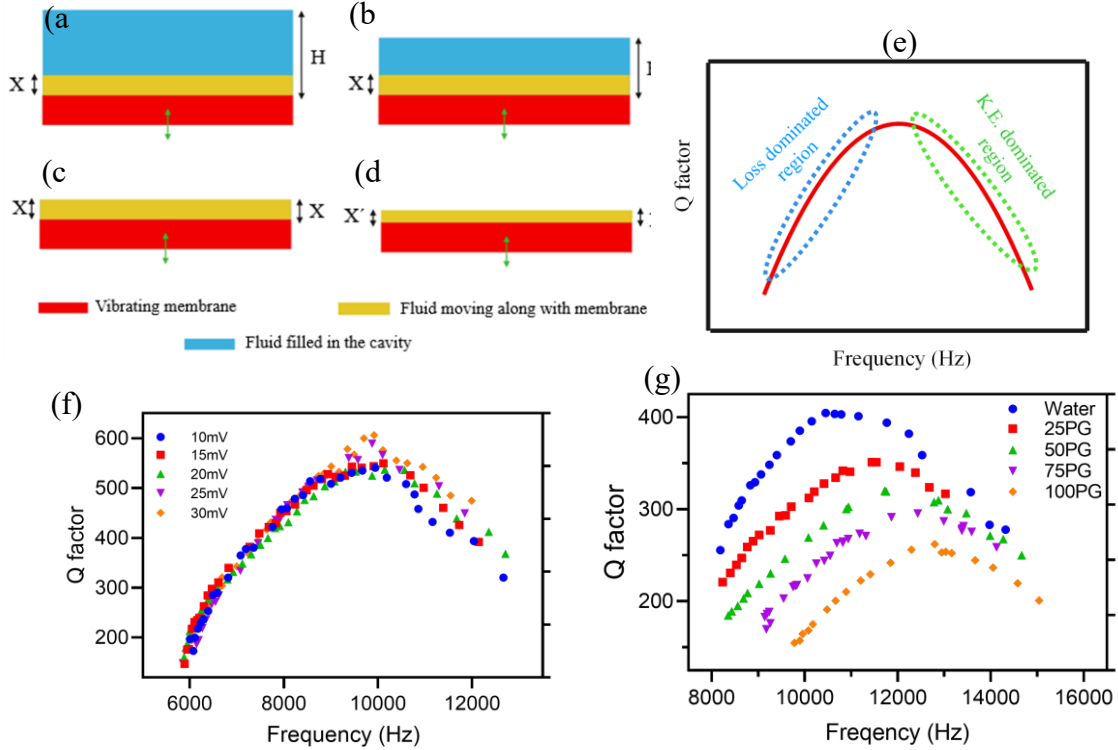

*Figure 5: The hypothesis explains the occurrence of the Q-enhancement as shown in the Figure. The RED indicates the vibrating membrane, BLUE (H & H`) indicates the fluid volume, and YELLOW (X & X`) represents the assumed volume that is actually vibrating with the membrane and the dependence of the $f_s$ on various parameters. (a)-(d) represents the scenarios at different instances while the fluid volume decreases; (e) The existence of a loss-dominated region and a K.E.-dominated region is shown and superimposed on the PMUT response as the fluid volume decreases. Figure not to scale. (f) Real-time response of the D4000 PMUT as the fluid decreases in the cavity at multiple actuating voltages. The co-linear path of graph in the beginning and diverging plots after $f_s$ can be seen. (g) Different mixtures of water and propylene glycol were pipetted into the D3500 PMUT cavity, and the response is shown in the graph. Increasing propylene glycol content shifts the curve rightward and downward.*

The arguments we have presented here are congruent with the experimental results and the simulation results. We have presented a reasonable explanation of the phenomenon we have observed. Clearly, a detailed analysis of the losses in the fluid medium immediately above the membrane and the air above the fluid column will be needed to establish a formula for determining the critical fluid volume that will maximize Q for a given resonator. It is evident that the losses depend on the viscosity of the fluid and hence the critical volume will also depend on the fluid properties, namely its density and viscosity.

There is one more non-obvious property that is also likely to play a role in determining this critical volume, and that is surface tension. While conducting these experiments, we noticed that the fluid evaporates rather quickly at low volumes. To further understand this dependence, we prepared fluid samples as mixtures of DI water ($\rho$= 990 kg/m$^3$ and $\eta$ = 1 cP) and propylene glycol (PG) ($\rho$= 1140 kg/m$^3$ and $\eta$= 34 cP) to vary the fluid properties. The resulting mixture's density and viscosity are listed in Supplementary Table 1. As seen in Supplementary Fig. 7, different mixtures of water and propylene glycol result in varied contact angles on silicon, ranging from 25.41° for 100% water to 15.03° for 100% propylene glycol. The surface tension of water in air is 72 mN/m and that of PG is 30 mN/m[45]. We carefully pipetted equal fluid volume from each of the mixtures into the PMUT cavity and monitored the dynamic response as the fluid mixtures evaporated. These experiments were performed in similar atmospheric conditions to eliminate any unintended effects of temperature and humidity variation. The PMUTs were actuated at a constant input voltage of 30 mV. The recorded responses are shown in Fig. 5(g). It is evident that with the increasing PG content, the amplitude of vibration at the $f_s$ reduces due to the increased viscosity. Additionally, the $f_s$ occurs at higher frequencies for fluids with increasing PG content. The change in the response in this case can be attributed to the combined effect of density, viscosity, and surface tension. To determine the role of density in the frequency shift, we filled a beaker with each fluid mixture, placed the PMUT at the bottom of the beaker (thus eliminating any significant role of surface tension), and measured the PMUT response. As see in Supplementary Fig. 8, the increasing density of the mixture causes a leftward shift in the resonant peak. However, we observe a rightward peak shift in Fig. 5(g) with increasing PG content. With such small volume of fluid, this shift can only be contributed to decreased surface tension (as measured by contact angles). Hence, the evaporation profile mediated by the surface tension has a role in dictating the frequency at which the maximum Q-enhancement occurs, along with viscosity that determines the amplitude of vibration at $f_s$. The role of surface tension on Q, however, particularly for the optimal fluid volume is not investigated here, and perhaps needs a more detailed study.

**Conclusions:**

In summary, our research presents a novel operational method for circular diaphragm-based micro-resonators, resulting in remarkably high Q-factors when used in fluids. We have extensively validated this phenomenon through finite element method simulations and meticulous experiments. Our circular diaphragm micro-resonator, specifically a PMUT, achieves the highest reported Q-factor in water (0.9 cP) at 602 in the fundamental vibrational mode, without any artificial device enhancements or gains. The exceptional combination of a high Q-factor, low operational frequencies (<10KHz), and inherent sensitivity positions PMUTs as valuable tools for early diagnosis and disease monitoring utilizing biological fluids like blood, serum, plasma, and CSF.

Our research further reveals that the Q-factor of a 2D circular micro-resonator, immersed in fluid on one side, is dependent on the fluid volume. Specifically, there exists an optimal fluid volume that enhances the Q-factor by at least an order of magnitude compared to air, thereby significantly increasing the sensitivity of the resonator for all sensing applications based on resonant peak shifts. The optimal fluid volume is influenced by fluid properties such as density, viscosity, and surface tension, although the exact relationship is yet to be determined.

Additionally, we have observed a consistent relationship between the height of the fluid column, corresponding to the optimal volume, and the resonator diameter (H/D≈0.03) for a given fluid. This discovery carries significant implications for sensing applications, particularly in disease diagnostics or health monitoring of biofluids, where small fluid volumes are often encountered. The one-order-of-magnitude Q-factor enhancement achieved simply through fluid volume offers a means to boost both sensitivity and the limit of detection without requiring other modifications.

In conclusion, our findings pave the way for the advancement of micro-resonators in sensing applications, presenting exciting possibilities for improving diagnostics and detection limits in various fields.

**Methods:**

The device shown in Fig. 1 was fabricated at the National Nano Fabrication Facility (NNFC), Indian Institute of Science, Bengaluru. While the detailed fabrication is described elsewhere[6], we provide a brief description of the process here. A PZT thin film of ~900 nm thickness is deposited on a platinized silicon-on-insulator (SOI) wafer with a 10 or 25 μm device layer. The wafer is patterned using photolithography to realize the top electrode. Further, the backside of the substrate is etched using DRIE to result in a membrane or a diaphragm. The PMUTs include three electrodes: a central electrode for actuation, an annular peripheral electrode for sensing, and a common ground electrode. The central electrode covers 70% of the PMUT area. The annular electrode is placed at a radial distance of 20 μm from the central electrode. To actuate the PMUT, an AC voltage of constant amplitude is applied across the central electrode while sweeping the frequency in the desired range. With an appropriate applied voltage, the membrane vibrates at its natural frequency resulting in the maximum amplitude motion. The resulting strain developed in the annular electrode gives rise to a voltage output measured across the annular (also called the *sense-electrode*) and the ground electrode. Several parameters such as diameter and thickness of the membrane, and the net residual stresses in the film stack can be engineered to arrive at a particular fundamental frequency of the PMUT structure. The PMUTs, after fabrication, are die bonded on a custom-made PCB and, subsequently, wire bonded using a TPT semiautomatic wire bonder. They are further bonded with epoxy to a 3D printed holder.

**Author Contributions:** S.H.P., A.R. and R.P., designed research; S.H.P. performed research; S.H.P., K.R., A.A., B.N., contributed fabrication/analyzing data; S.H.P., K.R. and R.P., wrote the paper.

**Acknowledgments**
The authors are grateful to Kritank Kalyan, Project Assistant (MEMS Lab) for Solidworks assistance and all the staff at the National Nanofabrication Centre (NNFC), Micro Nano Characterization Facility (MNCF) at the Centre for Nano Science and Engineering (CeNSE), IISc, the funding agencies – the Ministry of Human Resource and Development (MHRD), the Ministry of Electronics and Information Technology (MeitY).

**Competing Interests:**
The authors declare no conflict of interest.

**Data availability:**
All raw data is available with the authors and will be shared upon reasonable request. All the MATLAB codes used in the manuscript will be deposited in an online repository.

# Supplementary figures

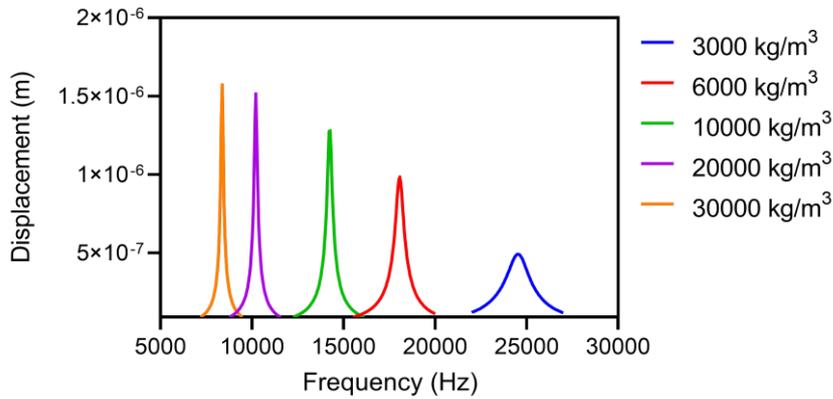

*Figure S1: COMSOL prediction of the displacements with increased density of the Si membrane. The model predicts the leftward shift of the resonant peak due to the increased mass and higher peak displacement due to the enhanced kinetic energy of the membrane.*

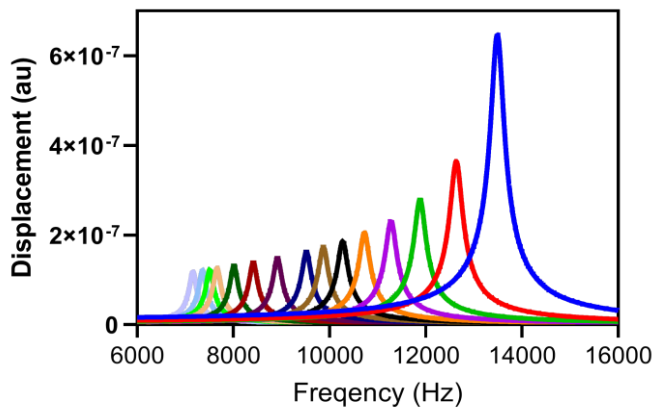

*Figure S2: Theoretical resonant curves with frequency matched to the fluid level in the cavity with a constant damping of ξ=0.01. As the fluid inside the cavity increases, the increased mass does not result in a enhancement of Q-factor thereby suggesting that damping has greater role to play as compared to the mass.*

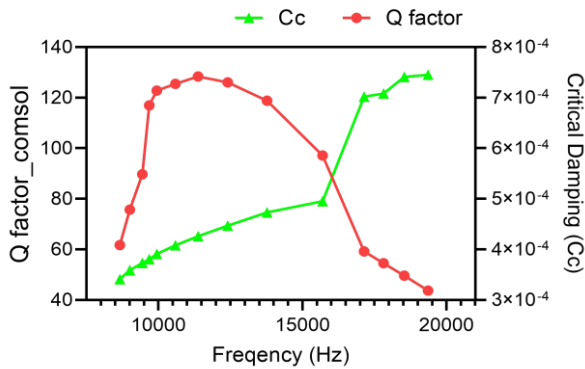

*Figure S3: Response of the critical damping co-efficient during the sweep spot. As seen in the figure, the $C_c$ is in a decreasing trend as the fluid level increases unlike the co-efficient of damping which passes through a minimum.*

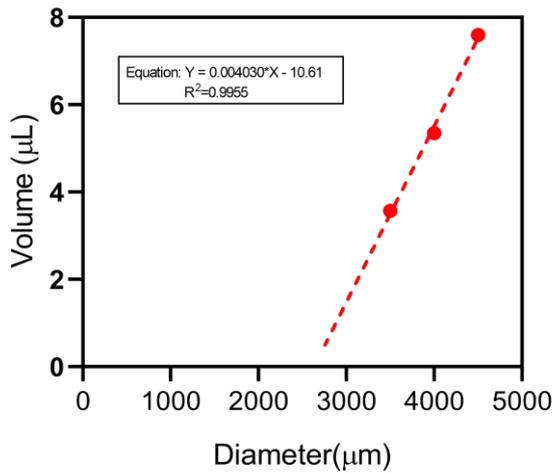

*Figure S4: Volume of water associated with special frequency at the maximum Q-factor is observed. These volumes have been calculated from the COMSOL data.*

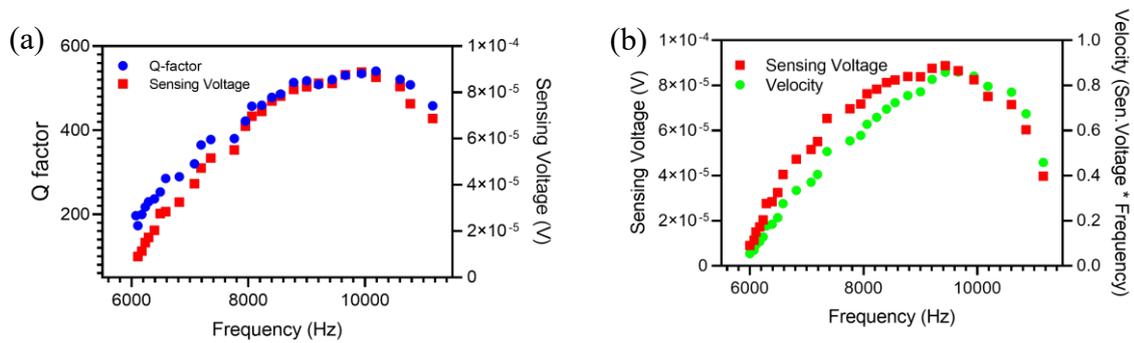

*Figure S5: The relation between the sensing voltage, Q factor and the K.E. of the device. (a) The actual output of the device is the sensing voltage (V), which is directly proportional to the displacement of the device. This displacement follows the same trend as the Q-factor around $f_s$. (b) The K.E. of the system (Displacement x Frequency) also follows the same trend as displacement and Q-factor and decreases after the $f_s$. This drop K.E. is due to the reducing fluid mass and the decreasing velocity.*

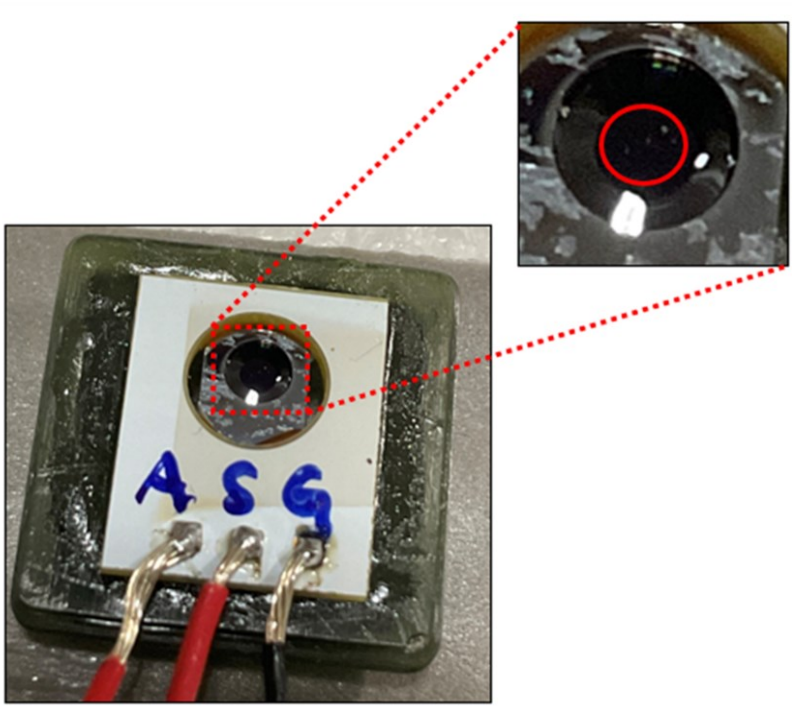

*Figure S6: As the fluid in the cavity evaporates, the added mass on the membranes decreases and the fluid film thickness reduces. After a threshold thickness, the fluid film is no longer stable due to the large displacements of the membrane causing spontaneous de-wetting. The figure shows an instant after spontaneous de-wetting occurs and inset shows the zoomed in image of the same. A boundary has been drawn to identify the de-wetted region.*

*Table S1: Fluid mixtures prepared by mixing water and propylene glycol (PG). The properties of the resultant fluid mixtures are given below. 25PG refers to 25% propylene glycol and 75% water*

| **Fluid Mixtures** | **Water** | **25PG** | **50PG** | **75PG** | **100PG** |
|---|---|---|---|---|---|
| **Density (Kg/m$^3$)** | 997 | 1007.25 | 1018.5 | 1029.25 | 1040 |
| **Viscosity (cP)** | 1 | 1.81 | 4.13 | 10.23 | 34.55 |

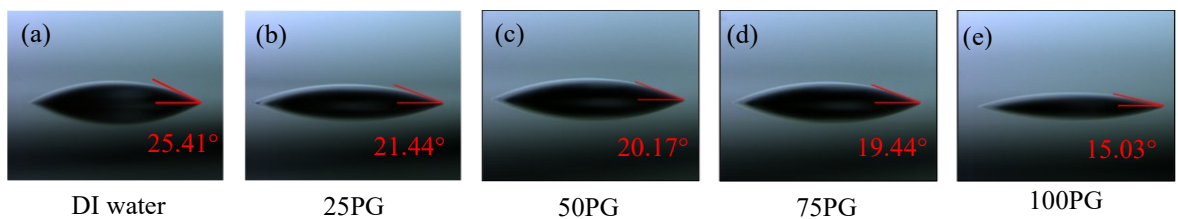

*Figure S7: Contact angle between the silicon surface and the fluid is measured. With increasing propylene glycol content, the contact angle decreases from 25.14° in DI water to 15.03° in 100% PG.*

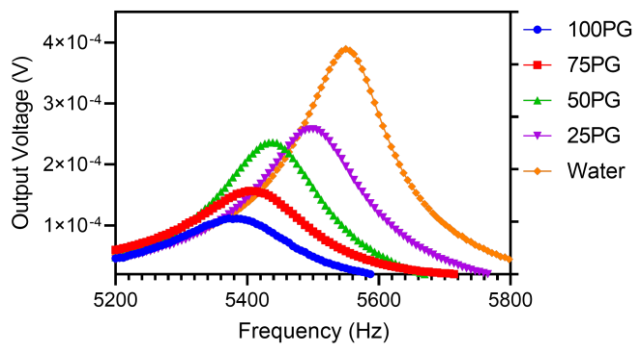

*Figure S8: Resonant response of a D3500 PMUT actuated at 30 mV under different fluid mixtures of water and PG. As the fluid mixture density increases from water to propylene glycol, the resonant curves shift leftward due to the increased added mass. The downward shift in the peak response is because of the increasing viscosity.*